# Developing an AI-Powered UX Research Point of View for Digital Health in A Regulatory Context: An Exemplar Case from MSM and Transgender HIV Care in Nigeria

**Emmanuel Oluwatosin Oluokun**
School of Computing and Engineering, Bournemouth University Poole, UK, eoluokun@bournemouth.ac.uk

**Festus Fatai Adedoyin**
School of Computing and Engineering, Bournemouth University Poole, UK, fadedoyin@bournemouth.ac.uk

**Huseyin Dogan**
School of Computing and Engineering, Bournemouth University Poole, UK, hdogan@bournemouth.ac.uk

**Nan Jiang**
School of Computing and Engineering, Bournemouth University Poole, UK, njiang@bournemouth.ac.uk

**Melike Akca**
School of Computing and Engineering, Bournemouth University Poole, UK, makca@bournemouth.ac.uk

**Abiodun Adedeji**
School of Computing and Engineering, Bournemouth University Poole, UK, adedejia@bournemouth.ac.uk

**Olumuyiwa Ayorinde**
School of Computing and Engineering, Bournemouth University Poole, UK, ayorindeo@bournemouth.ac.uk

**Fatima Ahmad Muazu**
School of Computing and Engineering, Bournemouth University Poole, UK, fahmad@bournemouth.ac.uk

## Abstract

User Experience Research (UXR) in a legal and regulatory contexts presents unique challenges that require specialised approaches to protect vulnerable populations whilst generating actionable insights. Digital consultation, appointment booking, and medication delivery platforms show promise for extending care access; however, their real-world effectiveness is curtailed by an absence of theoretically grounded user experience research (UXR) methodologies that adequately account for the psychosocial conditions of these populations. This paper introduces a Generative AI-augmented UXR methodology, grounded in the UXR Point of View (PoV) Playbook, to guide the design of psychologically safe, low-cognitive-load digital health interventions for MSM and transgender individuals living with HIV/AIDS in Nigeria. Drawing from empirical research involving co-design workshops, thematic analysis, and requirements engineering, the methodology is operationalised through a four-stage UXR process encompassing AI-supported hypothesis generation, foundational planning, insight generation via Building Blocks, and the construction of stakeholder-specific PoV narratives. This process results in ten theory-informed UXR Play Cards that translate psychological mechanisms and empirical findings into actionable design guidance. Each play contains actionable tasks, AI-augmented approaches, and ethical guardrails tailored for research with marginalised populations. The output is a set of ten theory-informed UXR Play Cards translating psychological insight and empirical evidence into actionable design guidance. The core contribution is a replicable, stigma-aware, and privacy-centred framework for responsible GenAI use in UXR practice, advancing human-centred digital health design for marginalised communities.

## 1. INTRODUCTION

Digital health interventions hold immense potential for improving health outcomes among marginalised populations, yet their design and implementation in vulnerable contexts present unprecedented UX research challenges [1]. In Nigeria, where the Same-Sex Marriage (Prohibition) Act 2014 criminalises same-sex relationships with penalties of up to 14 years imprisonment, men who have sex with men (MSM) and transgender people living with HIV/AIDS face compounded barriers to accessing antiretroviral therapy (ART) and health care services [2, 3]. HIV prevalence among MSM in Nigeria is estimated to be up to 19 times higher than in the general population, yet structural criminalisation suppresses the very care-seeking behaviours that digital health tools are designed to encourage [3].

User Experience Research in such contexts must navigate complex ethical terrain: how do we gather authentic user insights whilst protecting participants from legal persecution, social stigma, and physical harm? How can we design digital platforms that serve highly vulnerable populations without inadvertently exposing them?

The UXR PoV Playbook [4] provides a four-level pyramid framework i.e. Foundation, Data Collection, Insight Generation, and UXR PoV for distilling complex research data into actionable insights that drive product strategy and design. However, prior applications of this framework [5, 6] have focused on commercial or clinical contexts with accessible user populations. No existing adaptation explicitly maps research methods, plays, and outputs to each pyramid level when working with populations for whom research participation itself is dangerous. Existing UXR PoV frameworks do not adequately address the unique requirements of conducting research with marginalised populations, where every design decision carries potentially life-threatening consequences. Furthermore, the recent advancement in artificial intelligence capabilities offers new opportunities to enhance UX research rigour whilst maintaining participant safety through privacy-preserving technologies, intelligent anonymisation systems, and predictive risk assessment tools [6].

GenAI is employed throughout as a co-analytic tool, processing empirical evidence from a co-design study [7] and a systematic review [1] to accelerate synthesis, generate hypotheses, and articulate stakeholder-specific narratives under researcher-led oversight. The primary contribution of this work is a replicable, stigma-aware framework for responsible GenAI integration into UXR practice - one that advances human-centred approaches to digital health design for populations whose needs have been structurally marginalised in both healthcare systems and technology design.

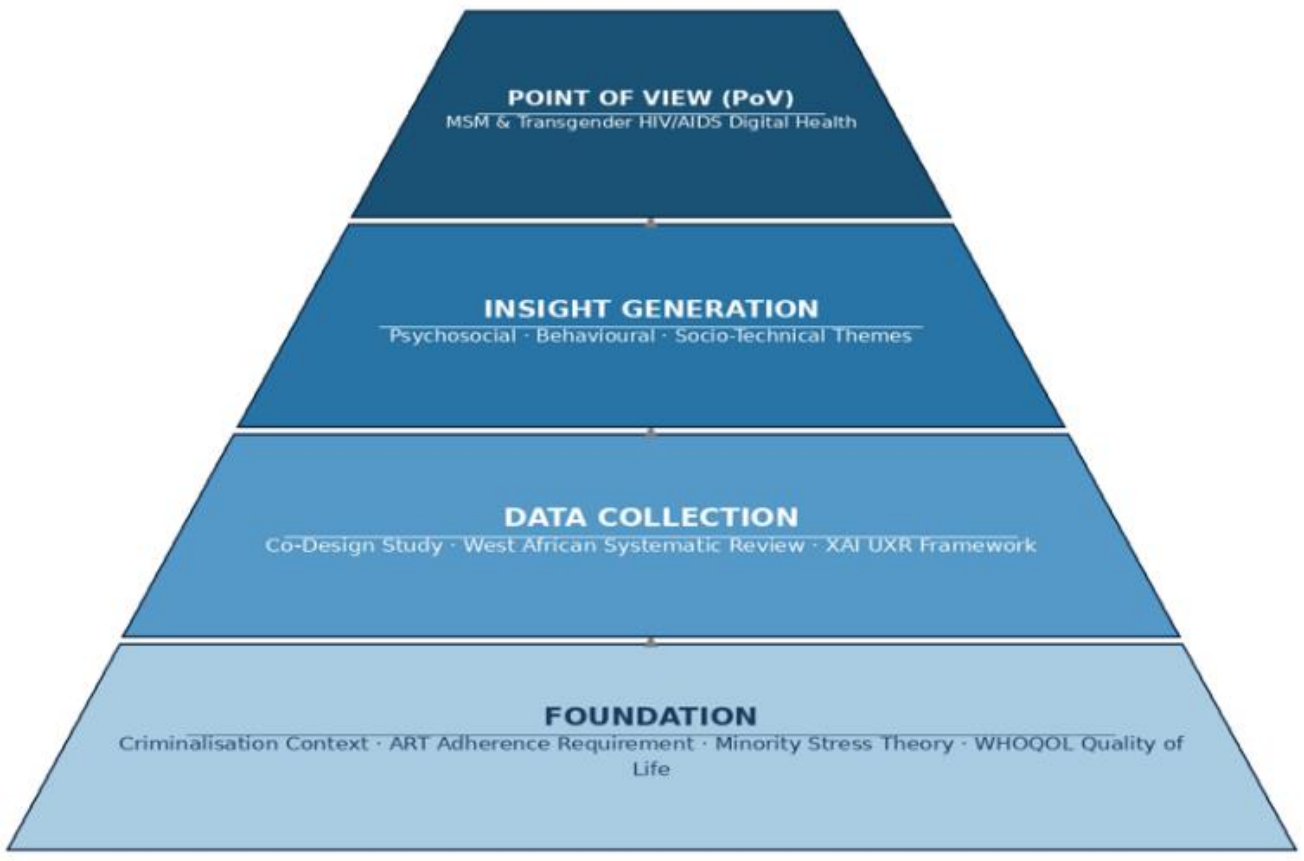

**Figure 1. UXR Point of View (PoV) Pyramid Framework**

*(The UXR Point of View (PoV) Pyramid Framework adapted for MSM and Transgender HIV/AIDS Digital Health, illustrating the four-tier progression from contextual foundation to actionable PoV)*

## 2. RELATED WORK

Research on digital health interventions for key populations (KPs) in marginalised settings has highlighted profound challenges in balancing research rigour with participant safety. Studies from sub-Saharan Africa have documented how criminalisation creates 'triple stigma'; HIV-related stigma, sexual minority stigma, and legal criminalisation that fundamentally shapes how individuals interact with health systems and digital technologies [3, 8]. Participants in such contexts express legitimate fears about data breaches, unintended disclosure, and the weaponisation of their health information by authorities or hostile actors [9].

Our systematic review of digital health interventions in West Africa [1] found that whilst mobile health (mHealth) solutions showed promise for improving ART adherence, most implementations failed to adequately address privacy concerns specific to underserved populations. SMS reminders containing explicit HIV related terminology risked unintended disclosure when phones were shared or accessed by family members, a common occurrence in resource-limited settings [1]. This underscores the need for UX research approaches that explicitly centre privacy, safety, and trust as foundational design requirements.

The UXR PoV Playbook [4] introduces a four-level pyramid framework i.e. Foundation, Data Collection, Insight Generation, and UXR PoV that guides practitioners in forming and communicating compelling points of view that drive product impact. The framework emphasises evidence-based insights, stakeholder alignment, and structured narratives that translate research findings into actionable design decisions. Recent applications have demonstrated its effectiveness in contexts [5, 6].

However, these applications have primarily focused on commercial contexts with willing, accessible user populations. The framework requires adaptation to address the unique ethical and methodological challenges of marginalised contexts, where traditional user research methods (e.g. focus groups, usability testing in controlled environments, collection of personally identifiable information) may expose participants to legal jeopardy.

Furthermore, the integration of AI tools offers opportunities to enhance research rigour whilst maintaining the anonymity and safety that marginalised populations require.

Recent scholars have explored how artificial intelligence can enhance various stages of UX research, from automated theme extraction and sentiment analysis to predictive modelling of user behaviour [6, 10]. Large language models (LLMs) and generative AI tools can assist researchers in synthesising vast amounts of qualitative data, identifying patterns across diverse user segments, and generating hypotheses for further investigation [4].

In the context of digital health for marginalised populations, AI offers promise for privacy-preserving research methods. Federated learning approaches allow predictive models to be trained on user data without centralising sensitive information [11]. Differential privacy techniques enable researchers to extract statistical insights from user behaviour whilst mathematically guaranteeing individual privacy [12]. Natural language processing can detect stigmatising or unsafe content before it reaches users, whilst conversational AI can provide health information and support without requiring users to disclose their HIV status to human providers. However, the deployment of AI in vulnerable populations also raises concerns about algorithmic bias, lack of transparency, and the potential for AI systems to perpetuate existing health inequities [13]. UX researchers must therefore approach AI integration with critical awareness, ensuring that technological capabilities serve user needs rather than simply optimising for technical metrics. This requires developing clear principles and practices for responsible AI deployment in marginalised health contexts.

## 3. METHOD

This study employed a GenAI-augmented evidence synthesis methodology, structured around the four-stage AI-powered UXR PoV framework [14]. The research integrates empirical data from two published primary sources with three theoretical lenses - Extended UTAUT, Minority Stress Theory, and WHOQOL - alongside Privacy Calculus and XAI Trust Theory, to generate design-actionable insights for digital health platforms serving MSM and transgender individuals with HIV/AIDS. Rather than applying a single behavioural model, the multi-framework approach was chosen to reflect the complexity of technology adoption in high-stigma, marginalised contexts, where psychological safety, infrastructure constraints, and institutional trust operate simultaneously as barriers to engagement. The four stages - (1) GenAI-supported evidence synthesis and hypothesis generation; (2) foundational stakeholder road mapping; (3) insight generation and Play Card development; and (4) PoV narrative construction - are illustrated in Figure 2.

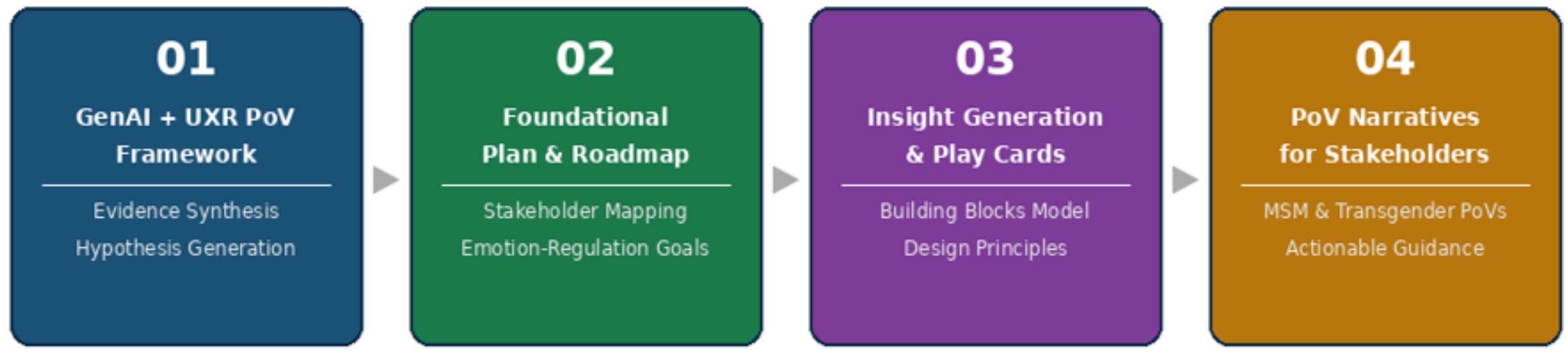

**Figure 2. Four-Stage AI-Powered UXR PoV Research Process**

### 3.1 Evidence Sources and Data Preparation

Two published empirical works formed the primary data corpus. The co-design study [7] employed participatory qualitative methods with clinicians, NGO staff, and care providers engaged with transgender communities in Nigeria. It documented psychosocial determinants of non-adherence, institutional challenges, and requirements for digital consultation and medication support. The West African systematic review of digital intervention [1] synthesised evidence from studies on digital health interventions for HIV/AIDS medication management and care delivery across West Africa, cataloguing phone-based adherence tools, teleconsultation systems, and their effectiveness in low-resource, high-stigma environments. A third source, an XAI UXR framework paper [6], supplied theoretical and empirical grounding on cognitive load, explainability, and AI trust calibration in health-adjacent digital systems.

All three documents were provided to the GenAI system alongside the UXR PoV framework paper [15] and PoV Playbook materials [14] to enable cross-document synthesis. Co-design data were structured thematically around user challenges, engagement patterns, and systemic barriers. Systematic review findings were organised by intervention modality, adherence outcome, infrastructure constraint, and confidentiality concern. This structured preparation ensured that AI-generated outputs were grounded in a well-organised evidence base rather than unstructured raw text.

### 3.2 Procedures and Prompts

**Stage 1: GenAI-Supported Evidence Synthesis and Hypothesis Generation**

The first stage aimed to extract design-relevant patterns from the evidence corpus and translate them into testable hypotheses linking UX variables to adherence and QoL outcomes. The AI was directed to surface recurring themes across psychosocial, behavioural, and design dimensions, cluster them within the UXR PoV analytical layers, and propose hypotheses grounded in the five theoretical frameworks. Prompts submitted to the GenAI system included:

- "Analyse the attached documents and identify recurring user challenges, engagement mechanisms, and accessibility barriers in enhancing the Quality of Life of Transgender and MSM individuals living with HIV/AIDS via Digital Consultation, Appointment Booking and Medication Delivery Platforms."
- "Cluster these insights into UX research themes that align with the UXR Point of View Framework (psychological → behavioural → design layers)."
- "Based on these synthesised themes, generate potential hypotheses and design opportunities relevant to emotion regulation, user engagement, and digital accessibility."

**Stage 2: Foundational Planning and Stakeholder Road mapping**

Stage 2 established the psychosocial and contextual grounding for the design process. GenAI was directed to contextualise how MSM and transgender individuals manage stigma and engage with digital tools, map the stakeholder landscape, and align design goals with the lived constraints of users in marginalised West African settings. Prompts submitted included:

- "Based on the systematic review and requirements analysis insights, explain how MSM and Transgender individuals living with HIV/AIDS typically manage emotions and stigmatisation, and what digital interventions can realistically support them without increasing cognitive overload."
- "Who are the primary and secondary stakeholders in a project developing digital tools for MSM and Transgender individuals living with HIV/AIDS? Describe their roles, motivations, and data needs."
- "Create a step-by-step project plan linking research goals, user needs, and stakeholder engagement following a mixed-method approach (quantitative, qualitative, AI synthesis)."

**Stage 3: Insight Generation and Play Card Development**

Stage 3 applied the Building Blocks model i.e. Foundation, Data Collection, Insight Generation, and PoV translation to produce ten UXR Play Cards. GenAI iteratively connected theoretical frameworks, empirical patterns, and design application, with researcher validation at each step. Prompts submitted included:

- "Cluster empirical data into themes of feedback, motivation, attention, and trust."
- "Generate hypotheses Play Card content linking theory, evidence, and design rationale."
- "Generate hypotheses Play Card content linking theory, evidence, and design rationale and use the example Play Cards as a reference when creating the cards."

**Stage 4: PoV Narrative Construction and Stakeholder Communication**

The final stage transformed Play Card insights into coherent, role-specific PoV narratives for MSM users, transgender users, clinicians, psychologists, NGOs, and design teams. GenAI supported iterative phrasing and thematic clustering; interpretive authority remained with the research team throughout. Prompts submitted included:

- "Check the UXR PoV template and generate MSM and Transgender-specific PoV narratives."
- “Generate App developers PoV narratives”
- "Refine the PoV statement for clinical accuracy and user experience clarity."
- "Ensure tone consistency with neurodiversity and therapeutic language."
- "Summarise insights from Play Cards into concise stakeholder messages."

## 4. RESULTS

The study yielded a structured progression of findings across the four UXR PoV stages: a theoretically grounded hypothesis set; a stakeholder-informed foundational plan; ten actionable UXR Play Cards; and population-specific PoV narratives for MSM and transgender users. Across all stages, Extended UTAUT, Minority Stress Theory, WHOQOL, Privacy Calculus, and XAI Trust Theory provided the theoretical architecture connecting empirical evidence to design guidance. The following sections report each stage in turn.

### 4.1 Synthesis and Hypothesis Generation

In the first stage, GenAI processed the two empirical sources and the XAI framework paper alongside the UXR PoV materials to surface recurring patterns and generate design-relevant hypotheses. Through iterative prompting and researcher-led validation, three analytical layers emerged i.e. psychosocial, behavioural, and socio-technical design. Each containing distinct but interconnected themes. The psychosocial layer encompassed stigma internalisation, fear of disclosure, identity conflict, and trust fragility. The behavioural layer captured missed doses, selective platform engagement, concealment-driven interaction patterns, and responsiveness to reinforcement. The design layer identified privacy-by-architecture, explainable AI, multi-channel access, and persona-sensitive pathways as the principal design requirements arising from these upstream dynamics.

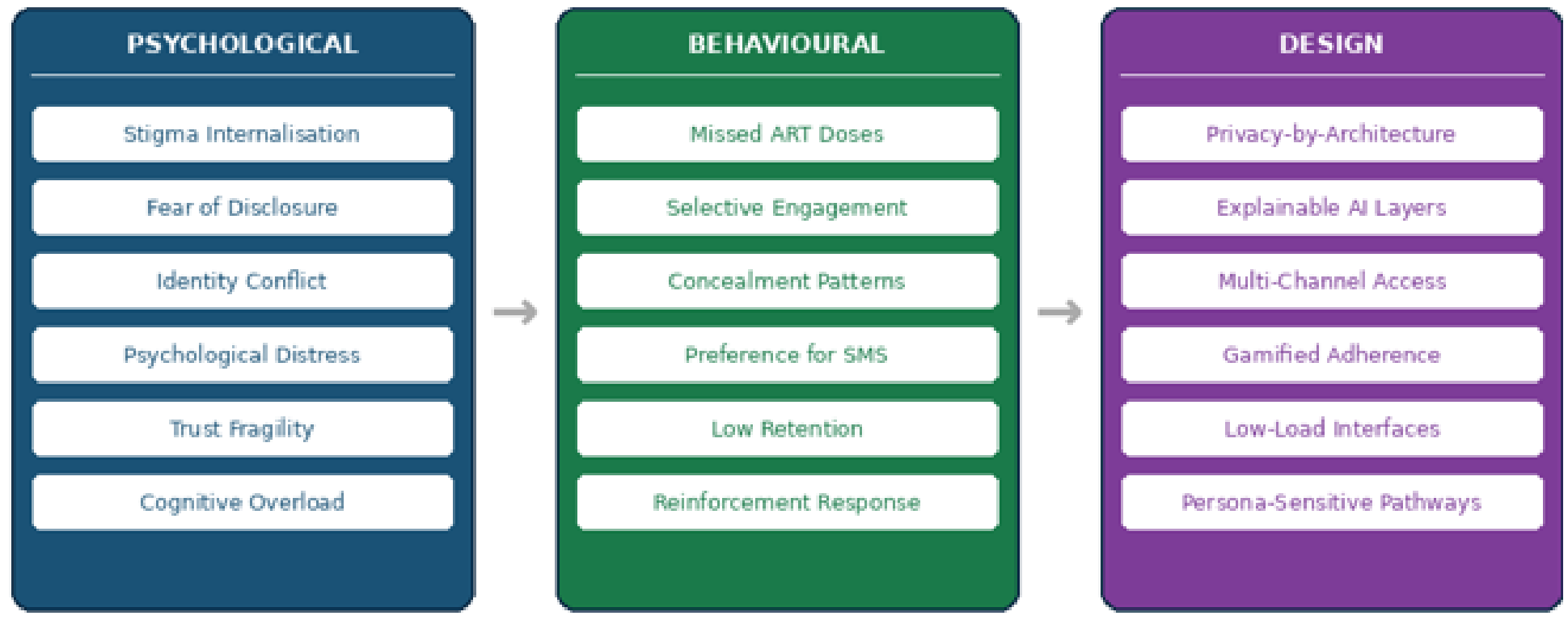

**Figure 3. Psychological → Behavioural → Design Translation Model**

This three-layer structure is illustrated in Figure 3 and formed the analytical backbone for all subsequent Play Card and PoV development. It illustrates how internal psychosocial states produce observable engagement patterns that must be addressed through deliberate design decisions. Ten hypotheses were co-generated from this synthesis. These are presented with their theoretical basis, evidence source, and design implication in Table 1.

**Table 1. GenAI-Generated Hypotheses with Theoretical Basis, Evidence Sources, and Design Implications**

| Hyp. | Statement | Theoretical Basis | Evidence Source | Design Implication |
|---|---|---|---|---|
| H1 | Perceived privacy control positively predicts emotional comfort during digital health interactions. | Extended UTAUT (Perceived Privacy); WHOQOL Environmental Safety | Disclosure fear as primary disengagement driver [1,7] | Privacy-by-architecture; discreet notification systems |
| H2 | Emotional comfort mediates the relationship between privacy perception and intention to use the platform. | Privacy Calculus; Minority Stress Theory | Confidentiality breach fear reduces adoption [1,7] | Stealth mode UI; customisable app naming |
| H3 | Lower perceived interface complexity increases sustained engagement among high-stigma users. | Cognitive Load Theory; XAI UXR Framework | Cognitive disengagement from complex AI systems [6] | One-task-per-screen; progressive disclosure |
| H4 | Facilitating conditions (connectivity, device access) moderate the relationship between usability and adoption. | Extended UTAUT (Facilitating Conditions) | Infrastructure barriers in West Africa [1] | SMS fallback; offline sync; low-data mode |
| H5 | Positive reinforcement feedback increases ART medication adherence consistency. | WHOQOL Psychological Wellbeing; Behavioural Habit Theory | Encouragement symbols and SMS reminders [7] | Private streak tracking; encouragement icons |
| H6 | Perceived clinician empathy positively predicts system trust. | Extended UTAUT (Social Influence); Trust Formation Theory | Teleconsultation integration improves adherence [7] | Secure in-app messaging; scheduled teletherapy |

| Hyp. | Statement | Theoretical Basis | Evidence Source | Design Implication |
|---|---|---|---|---|
| H7 | System trust predicts sustained platform engagement and QoL improvement. | Extended UTAUT (Behavioural Intention → Use); WHOQOL | Trust as central mediator for MSM/transgender users [1, 7] | Explainable AI; transparent data practices |
| H8 | Explanation clarity in AI-driven recommendations reduces cognitive disengagement. | XAI UXR Framework; Cognitive Bias Mitigation | Misinterpretation risk in AI outputs [6] | Layered explanation UI; confidence indicators |
| H9 | Neutral notification framing reduces anticipatory anxiety and increases reminder responsiveness. | Minority Stress Theory; Privacy Calculus | HIV-labelled SMS triggers avoidance [1] | User-defined phrasing; symbol-based alerts |
| H10 | Optional mood check-in features improve emotional awareness without increasing disengagement. | WHOQOL Psychological Domain; Cognitive Load Theory | ART fatigue and emotional distress patterns [7] | One-tap mood slider; no forced reporting |

### 4.2 Foundational Plan and Stakeholder Roadmap

In the second stage, GenAI was guided to synthesise the stakeholder landscape and map the psychosocial conditions shaping digital engagement for these populations. Drawing on the co-design study [7] and systematic review [1], the analysis identified eight stakeholder groups, each with distinct roles, needs, challenges, and data sensitivities.

Two structurally important tensions emerged from this mapping. First, a conflict between individual privacy and institutional reporting requirements: public health bodies and funders require outcome data that, if not carefully anonymised, risks re-identifying marginalised users. Second, a tension between emotional safety and technical feasibility: clinicians and developers prioritise functional completeness while users require minimal cognitive friction and maximum discretion. These tensions directly informed the Play Card development in Stage 3.

The mapping also confirmed the need to treat MSM and transgender users as distinct populations. MSM users primarily navigate stigma through the concealment of sexual identity, while transgender users contend with compounded discrimination arising from the intersection of HIV status and gender identity - a qualitatively different psychosocial profile demanding differentiated design responses. The complete stakeholder framework is presented in Table 2.

**Table 2. Stakeholder Mapping within the GenAI-Augmented UXR Framework for MSM and Transgender HIV/AIDS Digital Health**

| Stakeholder | Role | Primary Need | Key Challenge | Data Sensitivity |
|---|---|---|---|---|
| MSM Users | Core End-Users | Discreet, confidential ART support | Disclosure fear; identity concealment | Extremely Sensitive |
| Transgender Users | Core End-Users | Identity-affirming, integrated care | Intersectional stigma; device sharing | Extremely Sensitive |
| Clinicians | Supervisors/Advisors | Adherence data and tele-consult tools | Integrating remote care ethically | Clinical |
| Clinical Psychologists | Mental Health Support | Emotional distress indicators | Optional vs mandatory reporting | Emotional + Clinical |
| NGOs/CBOs | Community Intermediaries | Aggregated engagement metrics | Preserving anonymity in reporting | Aggregated |
| UX Researchers / Developers | Builders and Translators | Privacy-by-design and usability data | Balancing simplicity with clinical depth | Behavioural |
| Health Administrators | Operational Managers | Appointment and refill data | System integration and compliance | Operational |
| Ethics and Data Bodies | Governance Oversight | Consent and data minimisation protocols | Protecting marginalised populations | Compliance |

### 4.3 UXR Play Cards

Ten UXR Play Cards were developed through iterative GenAI synthesis and researcher validation in Stage 3. Each card operationalises a theoretical principle and empirical finding into a concrete, communicable design artefact. The cards are structured with a front face i.e. thematic title, issue type, and best practice; and a back face i.e. theoretical anchor, empirical evidence, testable hypotheses, and design play. This dual-face format, consistent with the UXR PoV Playbook structure [14], enables the cards to function as boundary objects bridging research evidence and interface design decisions.

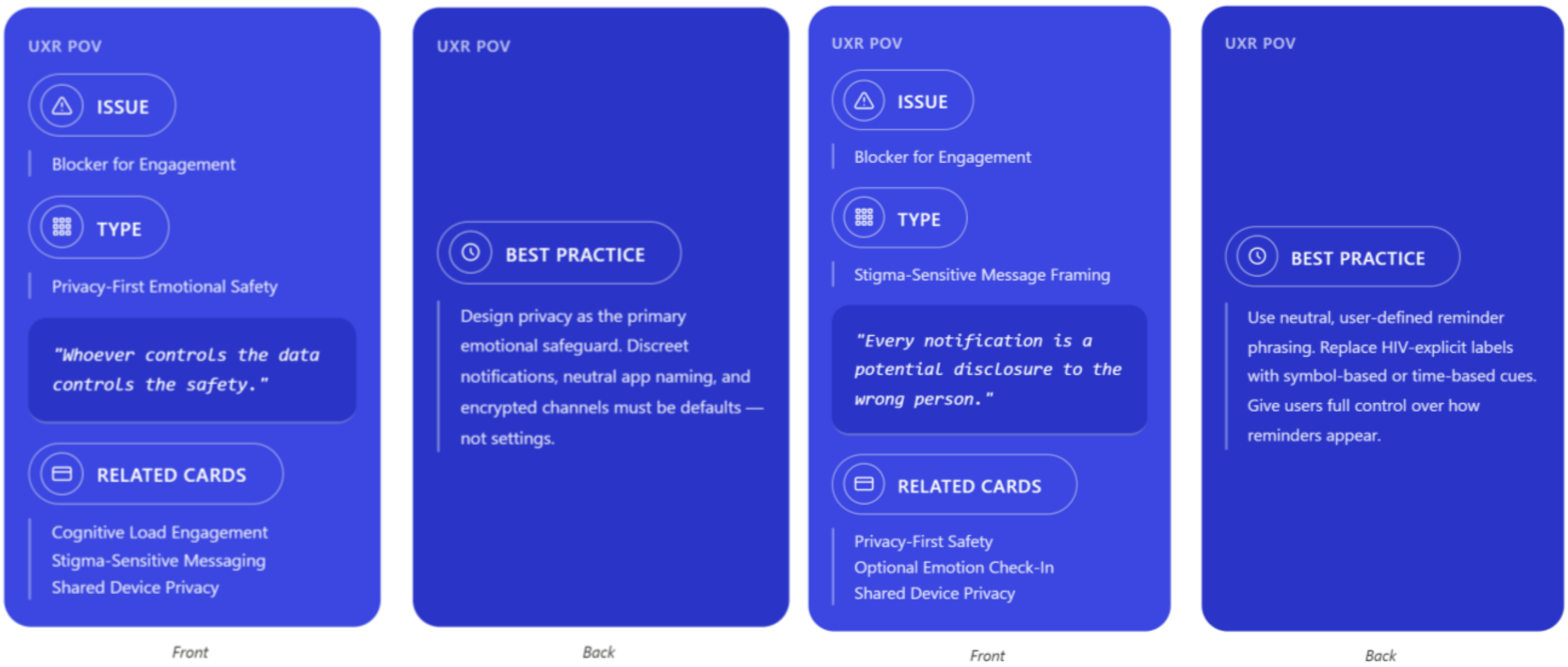

**Figure 4. UXR Play Card - Privacy Architecture** **Figure 5. UXR Play Card - Stigma Sensitive**

### 4.4 PoV Narratives

In the fourth stage, stakeholder-aligned PoV narratives were developed to operationalise the five theoretical frameworks into role-specific design strategies. Two population-specific PoV statements were refined for clinical precision and UX clarity through iterative researcher-AI collaboration.

> **MSM PoV Statement:** *For MSM living with HIV in high-stigma settings, digital health tools that support ART adherence and wellbeing are effective only when they demonstrably protect confidentiality, reduce cognitive and emotional load, and reinforce self-efficacy without amplifying stigma salience. In these environments, perceived psychological safety and user-controlled privacy are stronger determinants of sustained engagement than interface sophistication or feature breadth.*

> **Transgender PoV Statement:** *For transgender individuals living with HIV, digital consultation and health support tools must operate as identity-affirming, confidentiality-preserving care interfaces that coordinate ART adherence, psychological resilience, and gender-related health needs. Sustained participation depends on system trust, meaningful autonomy over data visibility, and frictionless interaction design - not the comprehensiveness of clinical information presented.*

> **App Developer PoV Statement:** *For app developers building digital health tools for MSM and transgender individuals living with HIV in marginalised settings, effective and ethical design is only achievable when privacy-by-architecture, low cognitive load, and stigma-sensitive interaction patterns are treated as non-negotiable technical requirements - not optional enhancements. In these contexts, the measure of a successful build is not feature completeness or interface elegance, but whether a vulnerable user can engage with the platform without fear of exposure, legal harm, or identity disclosure. Developers must resist the default pull toward transparency, social features, and data richness, recognising that the same design decisions that improve engagement in commercial apps can endanger lives in marginalised ones.*

Across all stakeholder groups, four cross-cutting design pillars emerged: psychological safety, cognitive simplicity, affirming identity architecture, and multi-channel accessibility. These provide a shared conceptual vocabulary for interdisciplinary teams working at the intersection of digital health, HCI, and HIV/AIDS care. The full PoV narrative framework is presented in Table 3.

**Table 3. GenAI-Supported PoV Narrative Development within the UXR PoV Framework for MSM and Transgender HIV/AIDS Digital Health**

| Stakeholder | Core Need | Primary Challenge | PoV Narrative Focus | GenAI Contribution |
|---|---|---|---|---|
| MSM Users | Confidential, discreet ART support | Disclosure risk and stigma-driven avoidance | Privacy-first design; discreet reminders; reinforcement-based adherence; low-load interaction | Iterative phrasing around stigma safety, autonomy, and habit scaffolding |
| Transgender Users | Identity-affirming, integrated care | Compounded stigma; identity invalidation | Affirming language; coordinated ART and gender care; optional psychological scaffolding | Synthesis of intersectional stigma patterns and affirming interface principles |
| Clinicians | Adherence visibility and teleconsulting tools | Integrating remote care into ethical clinical workflows | Explainable dashboards; secure messaging; discreet referral pathways | Structured feedback model synthesis and clinical transparency framing |
| Psychologists | Non-intrusive emotional support monitoring | Balancing therapeutic support with user autonomy | Voluntary distress tracking; one-tap crisis access; culturally sensitive prompts | Clustering of emotional regulation cues and psychosocial intervention triggers |
| NGOs / CBOs | Community-level engagement and trust | Maintaining privacy while capturing impact metrics | Stigma-sensitive design validation; aggregated reporting; community co-creation loops | Thematic grouping of community trust indicators with researcher validation |
| UX Researchers / Developers | Theory-to-interface translation with privacy-by-design | Absence of stigma-aware heuristics for key populations | Play Cards as shared design artefacts; privacy-by-architecture guidelines; low-bandwidth prototyping | Theory-to-interface mapping; adaptive logic articulation for vulnerable contexts |

## 5. DISCUSSION

This research explored how GenAI can be responsibly integrated into UXR practice to generate design guidance for digital health platforms serving and marginalised populations. The findings illustrate that GenAI can accelerate the synthesis of complex, multi-source evidence into structured design artefacts, but that this acceleration carries specific risks when applied to populations whose lived experience is characterised by structural vulnerability and psychosocial complexity.

The multi-framework theoretical architecture combining Extended UTAUT, Minority Stress Theory, WHOQOL, Privacy Calculus, and XAI Trust Theory proved essential for capturing the layered determinants of engagement in this context. Where earlier work on digital health adoption has applied single-framework models, the present study demonstrates that no single theory adequately explains why MSM and transgender individuals engage with or disengage from digital health tools. Perceived usefulness (UTAUT) interacts with disclosure risk (Privacy Calculus), which is itself modulated by minority stress exposure and its effect on anticipatory anxiety. System trust is simultaneously a function of clinician empathy (social influence), AI explainability (XAI Trust Theory), and infrastructure reliability (facilitating conditions). Quality of life outcomes are not simply downstream of behavioural intention but are co-produced by the emotional and social conditions under which platform interaction occurs. The extended structural model (Figure 3) captures these interdependencies and provides a quantitatively testable architecture for future validation.

GenAI demonstrated considerable value in several specific tasks: clustering qualitative co-design findings into thematic layers, identifying cross-document patterns across the systematic review and XAI framework, generating hypothesis sets that integrate multiple theoretical frameworks simultaneously, and articulating stakeholder-specific PoV narratives with consistent tone and structure. These capabilities align with the broader literature on GenAI as a cognitive collaborator in information-intensive research workflows [16]. However, the model exhibited a consistent limitation in calibrating the severity of criminalisation as a design variable. Several initial outputs adopted an implicitly optimistic framing recommending transparency features, community sharing, and social engagement mechanisms that would be actively dangerous for users operating in hostile legal environments. Researcher oversight was required to repeatedly reground AI outputs in the psychosocial realities documented in the co-design study [7].

This limitation has methodological significance beyond the present study. It suggests that GenAI training data, which skews toward majority-world, legally permissive contexts, systematically underrepresents the experience of marginalised and persecuted minorities. When deployed in UXR for such populations without reflexive researcher oversight, GenAI risks producing interface recommendations that embed majority-culture assumptions into platform design - a form of representational harm with concrete consequences for user safety [17]. The researcher-led, AI-assisted model adopted here in which GenAI scales and accelerates analysis while human researchers retain interpretive and validate authority represents a governance approach appropriate to this risk level.

The ten UXR Play Cards produced through this process advance the practical methodology of the UXR PoV framework [15, 14] in several ways. First, they operationalise multiple theoretical frameworks simultaneously within a single design artefact, rather than mapping each framework to a separate card. Second, they incorporate population-specific contextual constraints; shared devices, infrastructure inequality, criminalisation as first-order design considerations rather than edge cases. Third, they distinguish between MSM and transgender design requirements, resisting the homogenisation of key population needs that has characterised much digital health design to date. As boundary objects, these cards are designed to bridge the epistemic differences between clinicians prioritising therapeutic safety, developers prioritising technical feasibility, and NGOs prioritising community trust - a triangulation that the stakeholder mapping in Stage 2 identified as critical to platform legitimacy.

### 5.1 Limitations

Several limitations constrain the scope of these findings. The evidence corpus, while theoretically rich, was drawn primarily from Nigerian and West African contexts; the resulting design guidance may require adaptation for MSM and transgender populations in other legal, cultural, or infrastructural settings. The study employed a single GenAI system; future work should conduct cross-model comparisons to assess the consistency and potential biases of AI-generated insights, as demonstrated in comparable UXR PoV research [4]. Most critically, neither the Play Cards nor the PoV narratives produced here have been empirically validated with end users. The absence of direct participant involvement in this phase is a significant methodological limitation, given the centrality of participatory approaches to ethical design for marginalised communities. Prospective co-design workshops, usability studies, and field pilots are required before these artefacts can be considered validated design guidance.

## 6. CONCLUSION AND FUTURE WORK

This paper has presented a GenAI-augmented UXR methodology for designing digital health platforms that genuinely serve MSM and transgender individuals living with HIV/AIDS - not merely in a technical sense, but in a manner that is psychologically safe, identity-affirming, and appropriate to the structural realities of marginalised contexts. By anchoring the UXR PoV process in Extended UTAUT, Minority Stress Theory, WHOQOL, Privacy Calculus, and XAI Trust Theory, the study has produced a theoretically coherent and practically grounded set of design artefacts: ten UXR Play Cards, two population-specific PoV statements, a multi-framework structural model, and a stakeholder mapping framework.

The findings position psychological safety not as a supplementary design concern but as the foundational condition for any form of sustained engagement. Privacy architecture, low cognitive load, reinforcement-based habit formation, explainable AI, multi-channel access, and identity-affirming interface design are not independent features, they are interconnected design requirements arising from a shared psychosocial reality in which every platform interaction is evaluated against the risk of exposure, criminalisation, and harm.

Future work will pursue two primary directions. First, empirical validation through participatory co-design workshops with MSM and transgender users in Nigeria, using the Play Cards as shared design artefacts to ground community input in evidence-based principles. Second, integration of the UXR PoV outputs into the TechAids ontology (TechAids-Onto) to formalise capability engineering components for the TechAids platform. The research team welcomes collaboration with digital health practitioners, NGOs, policymakers, and community advocates to refine, adapt, and extend this framework across other contexts where marginalised populations require human-centred digital health design.